\documentclass[manuscript,screen,review，authoryear]{acmart}

\usepackage{amssymb}

\usepackage{amsmath,amsthm,amssymb,color, bbm}
\usepackage{enumerate}
\usepackage{mathtools}
\usepackage{array}
\usepackage{booktabs}
\usepackage{rotating}
\usepackage{tabularx}
\usepackage{multirow}
\usepackage{booktabs}
\usepackage{makecell}  
\usepackage{longtable}
\usepackage{threeparttable}

\usepackage{float}
\def\beq{\begin{equation}}
\def\eeq{\end{equation}}
\def\beqr{\begin{eqnarray}}
\def\eeqr{\end{eqnarray}}
\def\beqrs{\begin{eqnarray*}}
\def\eeqrs{\end{eqnarray*}}
\def\bet{\begin{theorem}}
\def\eet{\end{theorem}}
\def\bel{\begin{lemma}}
\def\eel{\end{lemma}}
\def\bg{\begin{figure}[tbph]\begin{center}}
	\def\eg{\end{center}\end{figure}}

\def\bc{\begin{center}}
\def\ec{\end{center}}
\usepackage{amsmath}
\usepackage{accents}
\usepackage{multirow}
\makeatletter
\def\widebar{\accentset{{\cc@style\underline{\mskip10mu}}}}
\def\Widebar{\accentset{{\cc@style\underline{\mskip8mu}}}}
\makeatother

\def\|{\Vert}

\def\boxit#1{\vbox{\hrule\hbox{\vrule\kern6pt\vbox{\kern6pt#1\kern6pt}\kern6pt\vrule}\hrule}}

\renewcommand\footnotetextcopyrightpermission[1]{}
\numberwithin{equation}{section}

{\newcommand{\killproofname}{\unskip\nopunct}}
{\newcommand{\killproofname}[1]{\unskip\aftergroup\ignorespaces\ignorespaces}}
\makeatother
\newcolumntype{Z}{>{\footnotesize\centering\arraybackslash}X}

\begin{document}

\title{An LLM-Powered Semantic Alignment Framework for Journal Recommendation}

\author{Yanglin Yan}
\authornotemark[1]
\email{2023310943@email.cufe.edu.cn}
\affiliation{%
 \institution{Central University of Finance and Economics}
 \city{Haidian}
 \state{Beijing}
 \country{China}}

\author{Zicheng Xie}
\authornotemark[1]
\email{2024211019@email.cufe.edu.cn}
\affiliation{%
 \institution{Central University of Finance and Economics}
 \city{Haidian}
 \state{Beijing}
 \country{China}}

\author{Tianchen Gao}
\authornotemark[1]
\email{gaotc@pku.edu.cn}
\affiliation{%
 \institution{Peking University}
 \city{Haidian}
 \state{Beijing}
 \country{China}}

\author{Rui Pan}
\authornotemark[2]
\email{ruipan@cufe.edu.cn}
\affiliation{%
 \institution{Central University of Finance and Economics}
 \city{Haidian}
 \state{Beijing}
 \country{China}}

\author{Hansheng Wang}
\email{hansheng@pku.edu.cn}
\affiliation{%
 \institution{Peking University}
 \city{Haidian}
 \state{Beijing}
 \country{China}}

\footnotetext[1]{These authors contributed equally to this work.}
\footnotetext[2]{Corresponding author.}

\renewcommand{\shortauthors}{Yan et al.}

\begin{abstract}
Journal recommendation is an important task in scholarly information systems. Existing approaches typically rely on supervised learning models, manually engineered features, or historical interaction data, which may limit their generalizability and interpretability. We propose an LLM-powered semantic alignment framework that formulates journal recommendation as a semantic matching problem between manuscript content and journal scope descriptions. The framework enables large language models (LLMs) to infer journal suitability directly from article titles, abstracts, keywords, and candidate journal information without task-specific training. Experiments are conducted using DeepSeek-V3 on a dataset of 23,609 articles from 49 journals in statistics and related fields. The proposed framework achieves Top-3, Top-5, and Top-10 accuracies of 40.23\%, 53.67\%, and 70.05\%, respectively. Additional analyses show that incorporating reference information generally improves recommendation performance and that recommendations remain highly stable across repeated runs, with an average Top-5 Jaccard similarity of 84\%. The framework also generates interpretable reasoning outputs that provide insights into the recommendation process. These findings demonstrate the potential of LLMs as a training-free and scalable paradigm for journal recommendation and scholarly decision support.

\end{abstract}


\received{20 February 2007}
\received[revised]{12 March 2009}
\received[accepted]{5 June 2009}

\maketitle
\newpage

\section{Introduction}
With the rapid expansion of academic journals and the increasing specialization of research fields, researchers face growing challenges in selecting the most appropriate journal for their work \citep{sugimoto2018measuring}. This process has become increasingly complex, as authors need to consider multiple factors, including journal scope, impact factor, readership, and publication speed \citep{rowley2022factors}. Submitting to an unsuitable journal not only reduces the chances of acceptance, but may also delay the dissemination of research findings. Consequently, journal selection remains an important and challenging task, particularly for early-career researchers who may be less familiar with journal scopes as well as experienced scholars entering new fields \citep{zhang2023scholarly}. In this context, effective journal recommendation systems are of growing interest. By automatically matching manuscripts with relevant journals, such systems may improve the efficiency and success rate of the submission process. They can also facilitate the diffusion of scientific knowledge and support the career development of researchers \citep{gundougan2023deep}. 

To address the growing complexity of journal selection, various journal recommendation systems have been developed. These systems evolve from basic text similarity tools to advanced deep learning models. Early systems such as eTBLAST \citep{errami2007etblast} and JANE \citep{schuemie2008jane} rely on TF-IDF-based similarity measures and recommend journals based on aggregated matching scores. Although effective for capturing lexical similarity, these methods are limited to surface-level text information. Later studies introduce machine learning and hybrid approaches. For example, \cite{wang2018content} propose a content-based recommender system for computer science publications using feature selection and softmax regression. In the biomedical domain, \cite{feng2019deep} develop Pubmender, a deep learning framework based on convolutional neural networks (CNNs) trained on biomedical article abstracts. More recently, transformer-based models have attracted increasing attention. For instance, \cite{gundougan2023deep} employ SBERT embeddings to capture semantic similarity between manuscripts and journal scopes, achieving improvements over classical baselines. Similarly, \cite{michail2023journal} explore transformer models trained on large-scale {\it Web of Science} data to identify latent patterns in journal selection. A broader review of scholarly recommendation systems is provided in \cite{zhang2023scholarly}. 

Journal recommendation is a specialized branch of recommender systems, which are widely adopted in areas such as e-commerce   \citep{alamdari2020systematic}, entertainment \citep{christensen2011entertainment}, social media \citep{guy2010social}, and academic publishing \citep{zhang2023scholarly}. Major publishers such as IEEE, Springer, and Elsevier develop journal suggestion tools to assist authors during the manuscript submission process. In academic research, publication recommendation systems have been extensively studied in computer science \citep{wang2018content}, biomedical research \citep{feng2019deep}, and computational linguistics \citep{ali2020paper}. Despite these advances, most existing journal recommendation systems remain fundamentally similarity-driven. They typically rely on predefined feature representations and fixed scoring functions. Consequently, these methods may have difficulty capturing the complex relationship between manuscript content and journal scope, particularly for interdisciplinary research. 

Recent progress in large language models (LLMs) provides a potential alternative by allowing manuscript information and journal descriptions to be evaluated jointly in natural language, rather than through similarity scores alone \citep{raffel2020exploring}. 
LLMs are particularly suitable for text-intensive tasks because they are designed to process, integrate, and summarize information expressed in natural language. In recent years, LLMs have been increasingly applied to scientific contexts, including literature review assistance \citep{tang2025large}, citation recommendation \citep{algaba2025large}, peer review support \citep{zhuang2025large}, and the analysis of scholarly documents \citep{mugaanyi2024evaluation}. These studies suggest that LLMs can effectively handle domain-specific terminology and complex scientific narratives when provided with appropriate contextual information. Journal recommendation naturally belongs to this class of tasks, since it mainly relies on textual descriptions of manuscripts and journals rather than explicit decision rules \citep{de2024explainable}. Motivated by these observations, this study explores the use of LLMs for journal recommendation through a prompt-based framework. We evaluate the recommendation performance of LLMs and further examine whether additional bibliometric context can improve recommendation accuracy.

\begin{figure}[!ht]
    \centering
    \includegraphics[width=1\linewidth]{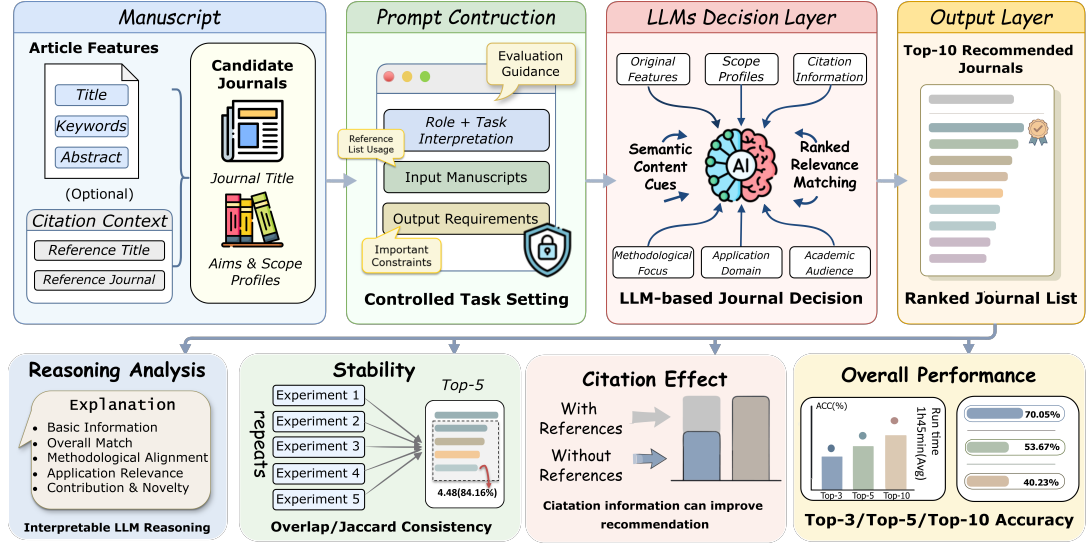}
    \caption{Overview of the proposed LLM-based journal recommendation framework. The framework takes manuscript information and candidate journal scope profiles as input, constructs structured prompts under controlled task settings, and uses LLMs to generate ranked journal recommendations.}
    \label{fig:overview}
\end{figure}

To this end, this study investigates an LLM-powered framework for journal recommendation. Our analysis is based on a dataset collected from the {\it Web of Science}, comprising 23,609 research articles published between 2021 and 2025 from 49 journals. These journals also constitute the candidate set for recommendation and mainly cover statistics and related fields. For each article, we collect structured bibliographic information and textual content, including title, abstract, keywords, and reference list. In addition, we collect the corresponding \textit{Aims \& Scope} statements from official journal sources to characterize the thematic focus of each candidate journal. Based on this dataset, we develop a unified LLM-powered prompting framework that formulates journal recommendation as a structured decision-making task in natural language. The overall workflow of the proposed method is illustrated in Figure~\ref{fig:overview}. 
Using this framework, we conduct a series of experiments to evaluate journal recommendation performance. These experiments include comparisons under different data settings (with and without reference list), accuracy analysis at different cutoff levels (Top-3, Top-5, and Top-10), stability analysis based on Top-5 overlap counts and Jaccard similarity, and additional investigations into potential information leakage and model reasoning behavior. 
Through these experiments, we aim to evaluate the effectiveness of LLM-based journal recommendation, examine the role of citation-related reference information in improving recommendation performance, and provide further insights into the relationship between model reasoning and recommendation outcomes.

The main contributions of this study are threefold. First, we provide an empirical evaluation of LLMs for journal recommendation based on a large-scale dataset of statistical publications. Specifically, we formulate journal recommendation as a natural language decision-making task, allowing the model to evaluate the alignment between manuscript content and journal scope directly from textual descriptions. Second, we propose a unified LLM-powered prompting framework that integrates textual content with reference-based information. In particular, we incorporate reference lists as additional contextual signals to enrich the representation of academic background information. This framework also enables comparisons under different information settings. 
Third, we conduct a series of experiments on recommendation accuracy, temporal stability, and reasoning behavior. The results show that incorporating the reference-based contextual information improves recommendation performance under most settings. We further observe from the reasoning analysis that accurate recommendations are often accompanied by clear and informative reasoning, whereas ambiguous reasoning may correspond to less precise ranking results. 
These findings provide useful insights for the design of AI-assisted journal recommendation systems.
Our recommendation system is now publicly available at \url{https://github.com/LLM-Journal-Rec/journal-recommendation-agent}. 

The remainder of this paper is organized as follows. Section~\ref{sec:Lit} reviews the related literature. In Section~\ref{sec:data}, we introduce our dataset and provide a descriptive analysis. The LLM-powered journal recommendation framework is introduced in Section~\ref{sec:prompt_design}, including the design of prompts from multiple perspectives. Section~\ref{sec:ACC} reports the main experimental results, including overall performance and the impact of incorporating citation information. The reliability, risks, and the reasoning behavior of the LLM-powered journal recommendation framework are examined in Section~\ref{sec:stability}.

\section{Literature Review}
\label{sec:Lit}
Journal recommendation is an important task in scholarly publishing. With the development of related technologies, a wide range of methods has been proposed to improve recommendation accuracy. This section reviews existing studies from several perspectives, including traditional recommendation methods, machine learning and deep learning approaches, and more recent frameworks based on Transformer architectures and LLMs.

\subsection{Traditional Journal Recommendation Approaches}

Early studies primarily adopt information retrieval-based approaches, including content similarity \citep{silva2015profile}, topic modeling-based similarity \citep{yang2012venue}, semantic structure similarity \citep{liu2019semantics}, and hybrid feature representations integrating multiple content aspects \citep{sakib2021hybrid}.
The first two approaches rely mainly on textual information, measuring similarity through word-level overlap or latent topic co-occurrence \citep{beel2016paper,amami2016lda}. In contrast, semantic structure-based methods explicitly incorporate domain knowledge to quantify semantic relatedness between papers. \cite{liu2017calculating}, for example, proposed a document-level semantic similarity framework that represents academic papers through structured topic events.
Building upon this semantic perspective, several studies further explored hybrid representations by integrating multiple content features into richer paper embeddings for recommendation tasks. Along this direction, \cite{kanakia2019scalable} combined co-citation and content similarity features to generate recommendations by leveraging heterogeneous information sources.

\subsection{Machine Learning and Deep Learning Approaches}

Subsequent studies introduced machine learning and deep learning methods to further improve recommendation performance.
At the text representation stage, manuscripts are encoded into feature vectors using techniques ranging from traditional weighted representations to distributed embeddings such as Word2Vec and Doc2Vec \citep{zhengwei2022recommendation}. 
Classical machine learning models, including gradient boosting trees \citep{collins2019meta} and support vector machines \citep{dehdarirad2020scholarly}, mainly learn decision boundaries for category prediction based on vectorized features.
\cite{zhengwei2022recommendation} employ Doc2Vec to embed titles, abstracts, and keywords, and subsequently train XGBoost to map these embeddings to journal categories.
With the increasing complexity of recommendation scenarios, neural network–based models, including CNNs \citep{kobs2020submit}, RNNs \citep{brewer2019personalized}, and GNNs \citep{shen2024temporal}, have demonstrated strong effectiveness in practical applications. Compared with traditional machine learning approaches, neural network–based models are more capable of capturing complex nonlinear patterns and hierarchical structures. \cite{feng2019deep} develop a deep learning–based recommender using CNNs to extract high-level representations from article abstracts. In a related direction, \cite{zhu2021recommending} adopt an attentive RNN model with heterogeneous knowledge embeddings for citation recommendation. More recently, \cite{iana2021graphconfrec} propose a GNN-based venue recommender that models relationships among papers, authors, and venues through graph structures.

\subsection{Recommender Systems Based on Transformers and LLMs}

In recent years, Transformer architectures have advanced natural language processing by capturing fine-grained and context-sensitive relationships among tokens.
Transformer-based models have become increasingly important in journal recommendation tasks, including BERT \citep{lee2018pre,macri2023evaluating} and its variants such as RoBERTa \citep{le2022simcpsr}, SciBERT \citep{thierry2023rar}, and DistilBERT \citep{zhang2022comparative}.
Among these approaches, \cite{reimers2019sentence} propose Sentence-BERT, which encodes text into high-dimensional vectors that can be efficiently compared using cosine similarity. 
In addition, \cite{haviana2024leveraging} combine BERT-based contextual embeddings with topic modeling to derive topic-level representations.

Following the emergence of Transformer techniques, LLMs have become a new breakthrough in recommender systems.
Recent studies mainly fall into two broad paradigms. The first treats LLMs as auxiliary components within existing recommendation frameworks \citep{boz2025improving}. Rather than replacing traditional methods, these approaches leverage the representational and reasoning capabilities of LLMs to enhance collaborative filtering and related recommendation models \citep{kim2024large,guan2026sear}.
The second paradigm places LLMs at the core of the recommendation system, focusing on designing specialized frameworks or reasoning mechanisms to strengthen recommendation performance. 
For example, \cite{bao2023tallrec} propose an instruction-tuning framework that aligns LLMs with recommendation tasks, substantially improving recommendation capability and cross-domain generalization.
Another line of research investigates whether LLMs are inherently capable of performing recommendation tasks. \cite{huang2026towards} construct a benchmark with diverse and realistic user queries to systematically evaluate LLMs as personalized recommendation assistants. At the user level, \cite{sun2025multi} evaluate LLMs as personalized movie recommenders through an online user study and found that LLMs exhibit strong explainability. In conversational settings, \cite{yun2025user} examine user experiences with LLM-based recommender systems through a diary study, highlighting their potential to support more personalized and exploratory recommendation interactions. 
However, the application of LLMs to academic venue recommendation remains largely underexplored, leaving open questions regarding how their semantic understanding and reasoning capabilities can be effectively utilized for this task.

\section{Data and Descriptive Analysis}
\label{sec:data}

In our primary study, we utilize a large-scale dataset derived from the {\it Web of Science}. The dataset contains 23,609 research articles published between 2021 and 2025 across 49 journals covering statistical theory (e.g., \textit{Annals of Statistics}), applied statistics (e.g., \textit{Annals of Applied Statistics}), computational statistics (e.g., \textit{Statistics and Computing}), and econometrics (e.g., \textit{Journal of Econometrics}). These journals serve as candidates for our recommendation system. A complete list of the candidate journals is provided in Table~\ref{tab:journals_full}.

\begingroup
\small
\setlength{\LTleft}{0pt}
\setlength{\LTright}{0pt}
\begin{longtable}{>{\raggedright\arraybackslash}p{0.58\linewidth}
                  >{\raggedright\arraybackslash}p{0.20\linewidth}
                  >{\raggedright\arraybackslash}p{0.16\linewidth}}
\caption{List of the 49 Candidate Journals, Their Abbreviations, and Categories}
\label{tab:journals_full} \\
\toprule
\textbf{Journal Name} & \textbf{Abbreviation} & \textbf{Category} \\
\midrule
\endfirsthead
\toprule
\textbf{Journal Name} & \textbf{Abbreviation} & \textbf{Category} \\
\midrule
\endhead
\bottomrule
\endfoot
ADVANCES IN DATA ANALYSIS AND CLASSIFICATION & ADAC & Data Science \\
AMERICAN STATISTICIAN & AS & Statistics \\
ANNALS OF APPLIED STATISTICS & AoAS & Applied Statistics \\
ANNALS OF STATISTICS & AoS & Statistics \\
ANNALS OF THE INSTITUTE OF STATISTICAL MATHEMATICS & AISM & Statistics \\
BERNOULLI & Bern & Prob. \& Statistics \\
BIOSTATISTICS & Biost & Biostatistics \\
BIOMETRICS & Bcs & Biostatistics \\
BIOMETRIKA & Bka & Statistics \\
COMMUNICATIONS IN STATISTICS-SIMULATION AND COMPUTATION & CSSC & Comput. Statistics \\
COMMUNICATIONS IN STATISTICS-THEORY AND METHODS & CSTM & Statistics \\
COMPUTATIONAL STATISTICS & CS & Comput. Statistics \\
COMPUTATIONAL STATISTICS \& DATA ANALYSIS & CSDA & Comput. Statistics \\
DATA MINING AND KNOWLEDGE DISCOVERY & DMKD & Data Science \\
ELECTRONIC JOURNAL OF STATISTICS & EJS & Statistics \\
JOURNAL OF APPLIED STATISTICS & JoAS & Applied Statistics \\
JOURNAL OF BUSINESS \& ECONOMIC STATISTICS & JBES & Econometrics \\
JOURNAL OF COMPUTATIONAL AND GRAPHICAL STATISTICS & JCGS & Comput. Statistics \\
JOURNAL OF ECONOMETRICS & JE & Econometrics \\
JOURNAL OF MACHINE LEARNING RESEARCH & JMLR & Machine Learning \\
JOURNAL OF MULTIVARIATE ANALYSIS & JMVA & Statistics \\
JOURNAL OF NONPARAMETRIC STATISTICS & JNS & Statistics \\
JOURNAL OF STATISTICAL COMPUTATION AND SIMULATION & JSCS & Comput. Statistics \\
JOURNAL OF STATISTICAL PLANNING AND INFERENCE & JSPI & Statistics \\
JOURNAL OF STATISTICAL SOFTWARE & JSS & Stat. Software \\
JOURNAL OF SURVEY STATISTICS AND METHODOLOGY & JSSM & Survey Statistics \\
JOURNAL OF THE AMERICAN STATISTICAL ASSOCIATION & JASA & Statistics \\
JOURNAL OF THE ROYAL STATISTICAL SOCIETY SERIES A-STATISTICS IN SOCIETY & JRSSA & Statistics in Society \\
JOURNAL OF THE ROYAL STATISTICAL SOCIETY SERIES B-STATISTICAL METHODOLOGY & JRSSB & Stat. Methodology \\
JOURNAL OF THE ROYAL STATISTICAL SOCIETY SERIES C-APPLIED STATISTICS & JRSSC & Applied Statistics \\
JOURNAL OF TIME SERIES ANALYSIS & JTSA & Time Series \\
MACHINE LEARNING & ML & Machine Learning \\
R JOURNAL & RJ & Stat. Software \\
SCANDINAVIAN JOURNAL OF STATISTICS & ScaJS & Statistics \\
SPATIAL STATISTICS & SpatStat & Spatial Statistics \\
STAT & STAT & Statistics \\
STATA JOURNAL & SJ & Stat. Software \\
STATISTICA SINICA & Sini & Statistics \\
STATISTICAL ANALYSIS AND DATA MINING & SADM & Data Science \\
STATISTICAL METHODS AND APPLICATIONS & SMA & Applied Statistics \\
STATISTICAL METHODS IN MEDICAL RESEARCH & SMMR & Medical Statistics \\
STATISTICAL MODELLING & SM & Stat. Modelling \\
STATISTICAL PAPERS & SP & Statistics \\
STATISTICS & Statistics & Statistics \\
STATISTICS \& PROBABILITY LETTERS & SPLet & Statistics \\
STATISTICS AND COMPUTING & SCmp & Comput. Statistics \\
STATISTICS IN MEDICINE & SMed & Medical Statistics \\
TECHNOMETRICS & Technometrics & Applied Statistics \\
TEST & TEST & Statistics \\
\end{longtable}
\endgroup

To further investigate whether citation information can improve recommendation performance, we additionally construct a citation dataset consisting of 2,683 articles published in eight core statistical journals selected from the 49 candidate journals. An overview of the key statistics for both datasets is shown in 
Table~\ref{tab:dataset-stats}. 
For each article, we collect five types of information, including title, abstract, keywords, reference titles, and reference journals, where the latter two are available only in the citation dataset. An illustrative example of the input features is presented in Table~\ref{tab:example}. In addition, for each candidate journal, we collect the corresponding \textit{Aims \& Scope} statement from the official journal website. These statements summarize the major research topics and application domains covered by each journal. 
A representative example is provided in Table~\ref{tab:example_scope}. To ensure consistency with the current \textit{Aims \& Scope} of the candidate journals, both datasets are restricted to recently published articles.

\begin{table}[ht]
\centering
\caption{An overview of the key statistics for both article datasets. Base dataset is 
used in the primary experiment (Section~\ref{sec:result_noref}); Citation dataset is used in the 
citation-augmented experiment (Section~\ref{sec:result_citation}).}
\label{tab:dataset-stats}
\begin{tabularx}{\linewidth}{
>{\raggedright\arraybackslash}l
>{\raggedright\arraybackslash}X
>{\raggedright\arraybackslash}X
}
\toprule
& \textbf{Base dataset} & \textbf{Citation dataset} \\
\midrule
% Source           & Web of Science & Web of Science \\
Period           & 2021--2025     & 2021--2023     \\
No.\ of Journals & 49             & 8              \\
No.\ of Articles & 23,609         & 2,683          \\
Fields           &  Title, Abstract, Keywords & Title, Abstract, Keywords, Reference titles, Reference journals \\
\bottomrule
\end{tabularx}
\end{table}

\begin{table}[htbp]
\caption{An illustrative example of an article in the dataset, drawn from a paper 
published in \textit{Biometrika}. In the first experiment, only the title, abstract, 
and keywords are used as input. The second experiment retains these fields and 
additionally incorporates the reference title and reference journal for each cited 
reference.}
\label{tab:example}
\renewcommand{\arraystretch}{1.15}
%\small
\begin{tabularx}{\linewidth}{
>{\raggedright\arraybackslash}p{2.2cm}
>{\raggedright\arraybackslash}X
}
\toprule
\textbf{Variable} & \textbf{Example} \\
\midrule
% Publisher & BIOMETRIKA \\
Title & A minimum aberration-type criterion for selecting space-filling designs \\
Abstract & Space-filling designs are widely used in computer experiments. Inspired by the stratified orthogonality of strong orthogonal arrays, we propose a criterion of minimum aberration-type for assessing the space-filling properties of designs based on design stratification properties on various grids. A space-filling hierarchy principle is proposed as a basic assumption of the criterion. The new criterion provides a systematic way of classifying and ranking space-filling designs, including various types of strong orthogonal arrays and Latin hypercube designs. Theoretical results and examples are presented to show that strong orthogonal arrays of maximum strength are favourable under the proposed criterion. For strong orthogonal arrays of the same strength, the space-filling criterion can further rank them based on their space-filling patterns.\\
Keywords & Computer experiment, Generalized minimum aberration, Space-filling hierarchy principle, Space-filling pattern, Strong orthogonal array \\
% Year & 2021 \\
\midrule

References
& [1] Optimal Sliced Latin Hypercube Designs, \textbf{Technometrics}. \\
& [2] Orthogonal and nearly orthogonal designs for computer experiments, \textbf{Biometrika}. \\
& [3] Analysis Methods for Computer Experiments: How to Assess and What Counts?, \textbf{Statistical Science}. \\
& [4] Generalized resolution and minimum aberration criteria for Plackett--Burman and other nonregular factorial designs, \textbf{Statistica Sinica}. \\
& [5] \dots \\

\bottomrule
\end{tabularx}
\end{table}

\begin{table}[htbp]
\caption{An example of a journal scope from Biometrika, obtained from the official website.}
\label{tab:example_scope}
\renewcommand{\arraystretch}{1.2}
\begin{tabular}{p{3cm} p{11cm}}
\hline
\textbf{Journal} & \textbf{Scope} \\
\hline
BIOMETRIKA
 &The three primary criteria by which all papers are judged are:
 
1.	importance of the work to the understanding, practice or potential practice of statistics;

2.	the degree of conceptual novelty and insight;

3.	the conciseness and clarity with which the ideas are conveyed.

The novelty may take many forms, including: the development of new methodology accompanied by an insightful analysis; papers where the innovation is around computational efficiency, where this is important for the application of the statistical method; a penetrating exposition of anomalous or unforeseen behaviour of mainstream inferential tools; original formulations uncovering foundational structure with potential relevance to statistical practice. Papers concerned purely with sampling properties of existing procedures or minor developments thereof are typically unsuitable unless such properties reveal considerable structural understanding. We do not publish purely applied work.

Biometrika publishes three types of paper: regular papers, synthesis papers (both normally fewer than 20 pages) and miscellanea articles (max. 8 pages). Miscellanea papers are not expected to have the breadth of contribution of regular papers but need to meet the same standards in terms of importance of the work, novelty of the insights, and general quality. Synthesis papers should either open up an emerging area outside of statistics to a statistical audience, assuming that area will be important to statistics, or be a synthesis of two or more areas of statistics that brings fundamental new insight.
 \\

\hline
\end{tabular}
\end{table}

\begin{figure}[h]
    \centering
    \includegraphics[width=1\linewidth]{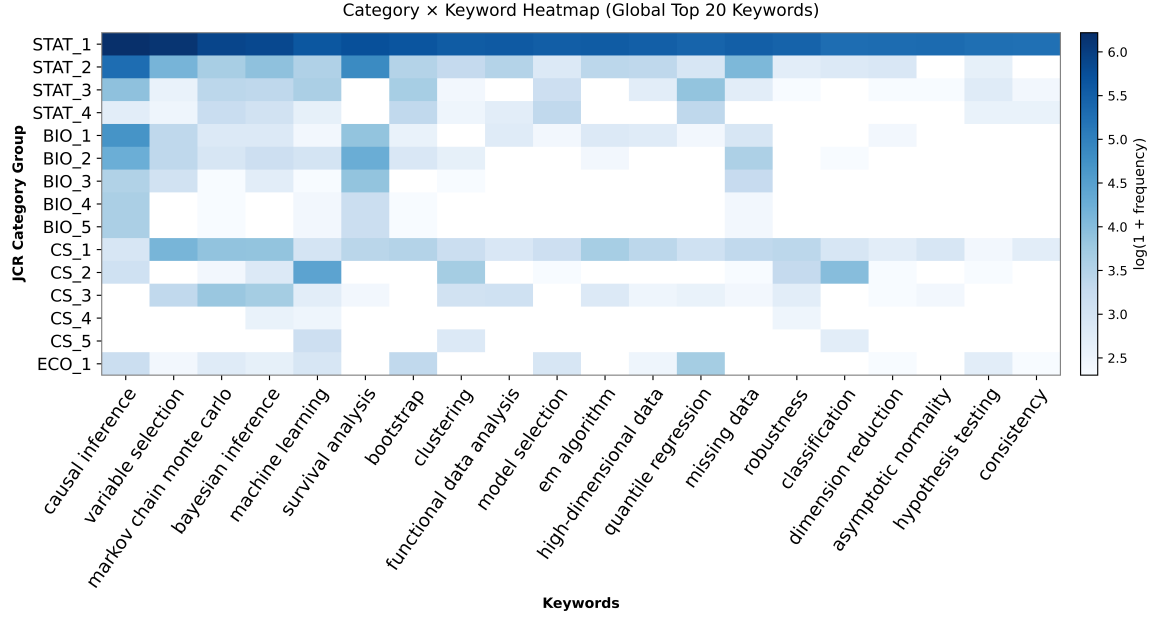}
    \caption{Frequency distribution of the top-20 keywords across aggregated JCR category groups. Original JCR categories are combined into broader disciplinary groups, including Statistics (STAT), Biology \& Medicine (BIO), and Computer Science (CS), where numeric suffixes denote subcategories within each group (e.g., STAT\_1 and STAT\_2). Rows are ordered by disciplinary group and the total keyword frequency within each group. The complete mapping between abbreviated labels and the original JCR categories is provided in Table~\ref{tab:category_mapping}.
}
    \label{fig:bubble}
\end{figure}

To examine whether article metadata can facilitate LLM-based journal recommendation, we conduct a descriptive analysis of author-provided keywords. Specifically, we investigate the relationship between keyword frequencies and journal characteristics, including JCR categories and individual journals. 
Given the large number of articles in our dataset, we focus on the top-20 most frequent author-provided keywords. The corresponding JCR categories are obtained from the \textit{Web of Science} based on the publishing journals. When a journal belongs to multiple JCR categories, only its primary category is retained for analysis.
We first investigate the relationship between the frequencies of the top-20 keywords and JCR categories.
As shown in Figure \ref{fig:bubble}, three main observations emerge. First, statistics-related categories exhibit a broad and dominant presence across nearly all keywords, highlighting their foundational role in methodological research. 
Second, biomedical categories display a more concentrated pattern, with relatively high frequencies for keywords related to causal inference, survival analysis, and missing data, reflecting the methodological specificity of biomedical applications. 
Third, computer science-related categories show higher frequencies for machine learning, clustering, and classification, emphasizing a stronger focus on algorithmic and scalable data-driven approaches.

\begin{table}[htbp]
%\footnotesize
\caption{Mapping of grouped labels to original JCR categories. This table reports the mapping between the abbreviated labels used in Fig.~\ref{fig:bubble} and the corresponding original JCR categories.}
\label{tab:category_mapping}
\begin{tabularx}{\linewidth}{l l X}
\toprule
\textbf{Group} & \textbf{Label} & \textbf{Original JCR Category} \\
\midrule

\multirow{4}{*}{STAT} 
& STAT\_1 & Statistics \& Probability \\
& STAT\_2 & Mathematical Methods \\
& STAT\_3 & Mathematics, Interdisciplinary Applications \\
& STAT\_4 & Social Sciences, Mathematical Methods \\
\midrule

\multirow{4}{*}{BIO} 
& BIO\_1 & Biology \\
& BIO\_2 & Medical Informatics \\
& BIO\_3 & Public, Environmental \& Occupational Health \\
& BIO\_4 & Health Care Sciences \& Services \\
\midrule

\multirow{4}{*}{CS} 
& CS\_1 & Computer Science, Artificial Intelligence \\
& CS\_2 & Computer Science, Information Systems \\
& CS\_3 & Computer Science, Interdisciplinary Applications \\
& CS\_4 & Automation \& Control Systems \\
\midrule

\multirow{1}{*}{ECO} 
& ECO\_1 & Economics \\
\midrule

\multirow{2}{*}{GEO} 
& GEO\_1 & Geosciences, Multidisciplinary \\
& GEO\_2 & Remote Sensing \\
\bottomrule
\end{tabularx}
\end{table}

\begin{figure}[h]
    \centering
    \includegraphics[width=1\linewidth]{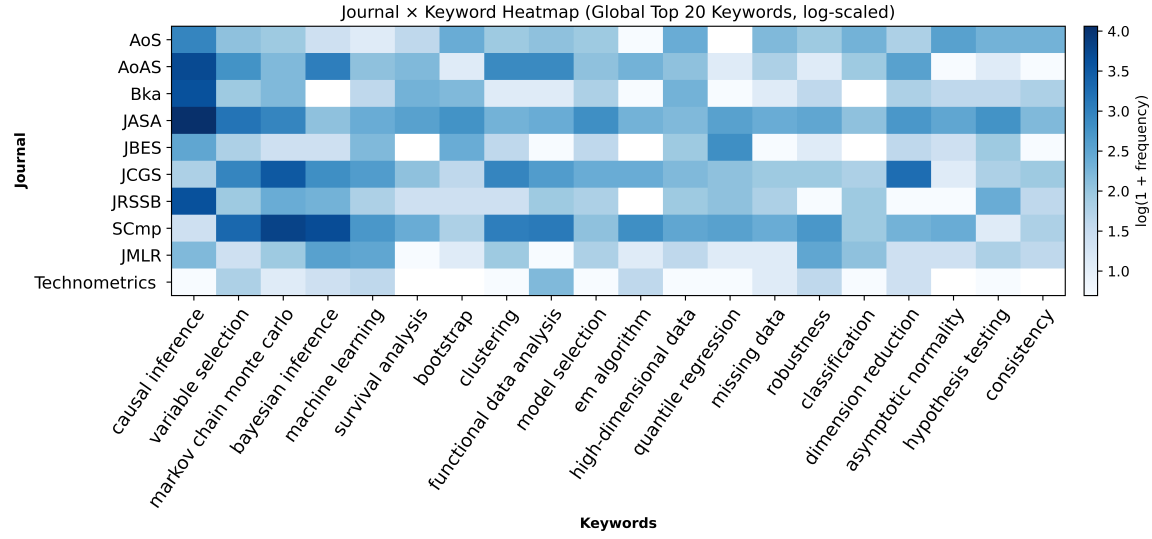}
    \caption{Frequency distribution of the top-20 keywords across the selected 10 journals. These journals similarly span the three major domains of statistics, biomedicine, and computer science.}
    \label{fig:bubble_1}
\end{figure}

We next examine the relationship between keyword frequency and individual journals while retaining the same set of top-20 keywords. To ensure representativeness, we select 10 journals based on impact factor and diversity in research scope.
As shown in Figure \ref{fig:bubble_1}, two main observations are discussed below. First, the selected keywords are distributed unevenly across journals. For example, keywords related to causal inference are more concentrated in journals such as \emph{AoAS} and \emph{JASA}, whereas keywords associated with Markov Chain Monte Carlo appear more frequently in journals such as \emph{JCGS} and \emph{SCmp}. Second, journals with a stronger theoretical orientation (e.g., \emph{AoS} and \emph{JASA}) exhibit higher frequencies of keywords related to causal inference, variable selection, and Bayesian inference, while journals with a stronger computational focus (e.g., \emph{JCGS}) are more strongly associated with clustering, high-dimensional data, and algorithmic modeling. Across both analyses, keyword distributions exhibit clear heterogeneity across JCR categories and individual journals, suggesting that article features contain informative signals for distinguishing research focus and supporting journal recommendation. 

\section{Prompt Design and Implementation}\label{sec:prompt_design}

The prompt plays a central role in the recommendation process, as its design largely determines whether the model can effectively understand the task and utilize the input information.
To adapt the journal recommendation task to LLMs, we design a prompt framework consisting of three instructional parts: main task  specification, input description, and inference with output formatting. Within this framework, the model is required to select and rank the most suitable journals from a predefined candidate set based on the article information provided. In addition, the practical implementation of the proposed recommendation system is demonstrated at the end of this section.

\begin{figure}[h]
    \centering
    \includegraphics[width=1\linewidth]{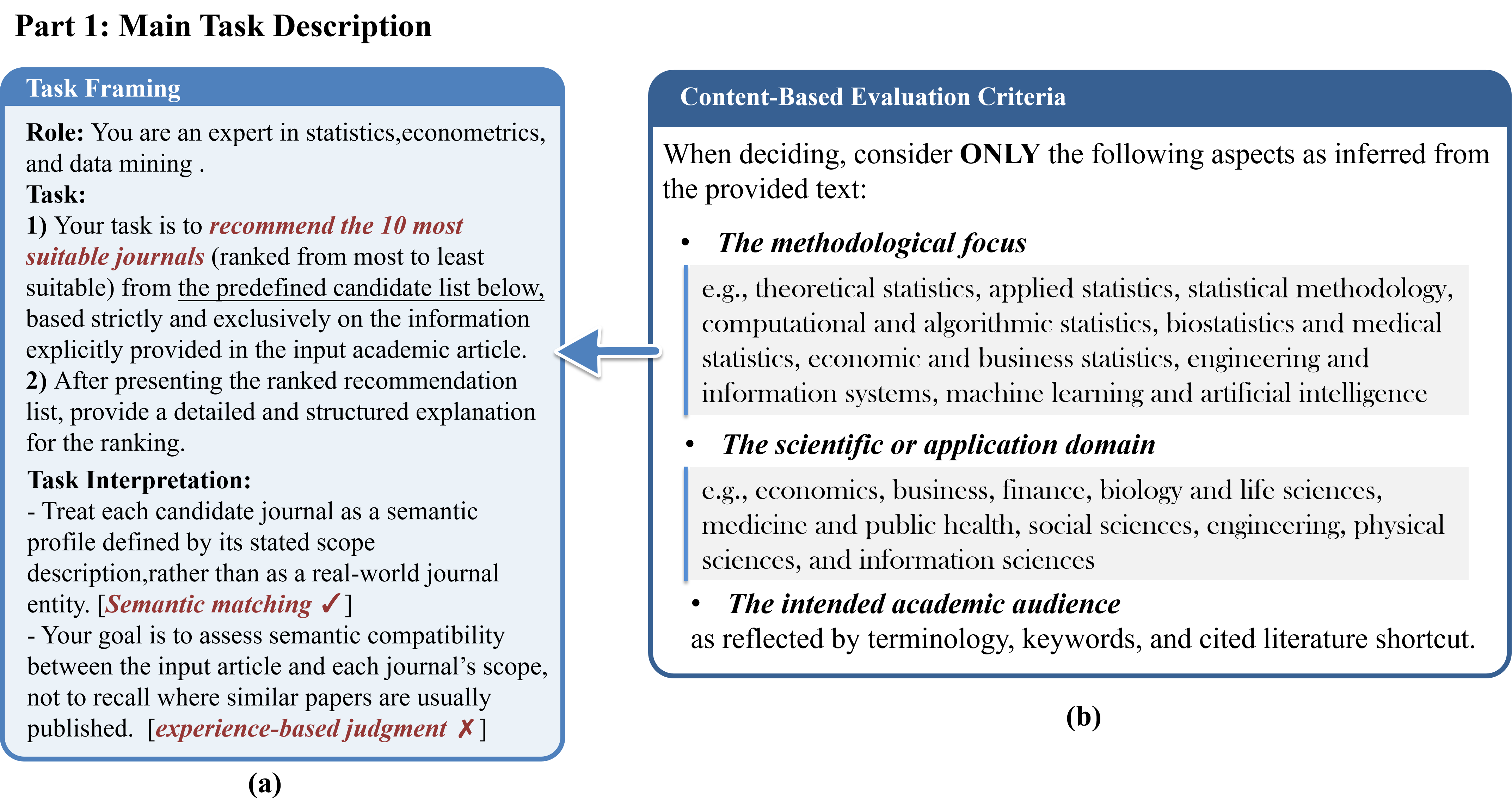}
    \caption{Illustration of the first component of the proposed prompt framework. Panel (a) presents the task framing through role assignment and task interpretation, while panel (b) defines three content-based evaluation criteria used to guide the journal recommendation process.}
    \label{fig:prompt_1}
\end{figure}

\paragraph{Part 1: Main Task Description} The prompt begins by defining the model as an expert in statistics, econometrics, and data mining, as shown in Figure~\ref{fig:prompt_1}. 
The main task then instructs the model to recommend the 10 most suitable journals for 
a given article based solely on the information provided in the input, ranked from 
most to least suitable, and to provide a structured explanation for each recommendation.
In our prompt design, the task is framed as evaluating the compatibility between the article content and the scope of each candidate journal. To achieve this, the prompt specifies three criteria for matching articles to journals: the methodological focus of the study, the scientific or application domain, and the intended academic audience as reflected in the terminology and keywords used in the article.

\begin{figure}[h]
    \centering
    \includegraphics[width=1\linewidth]{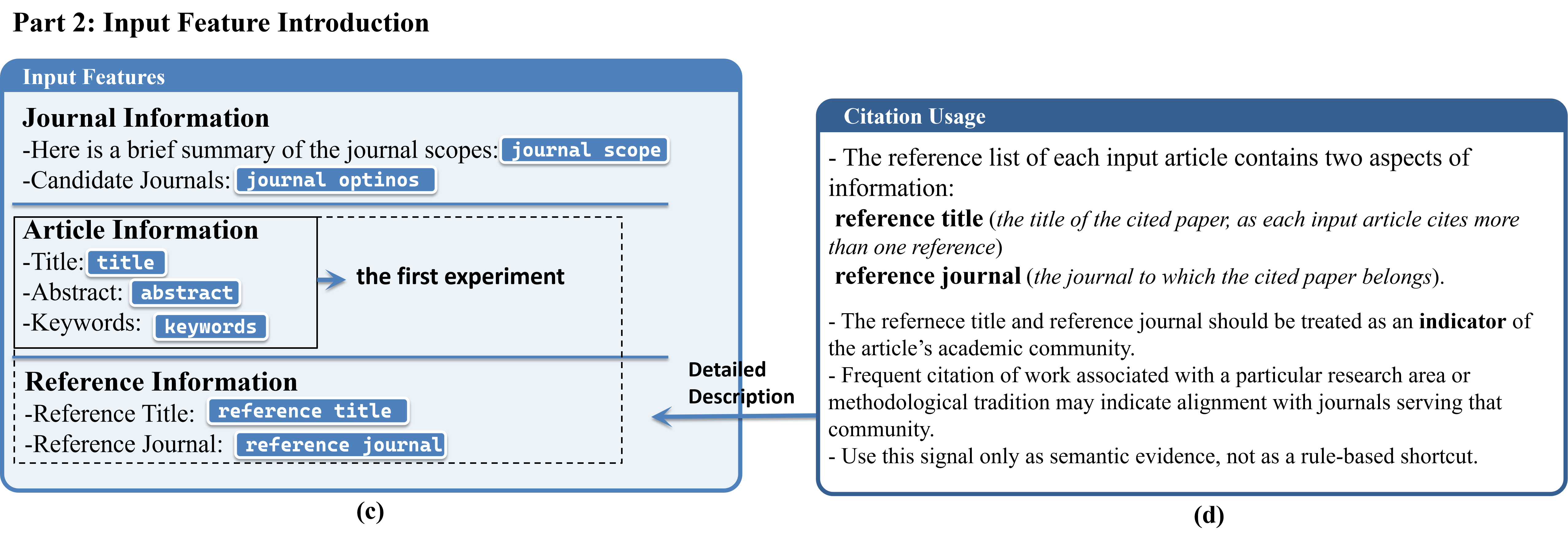}
    \caption{Illustration of the second part of the proposed prompt, covering the input description. Panel (c) presents the input features including journal scope descriptions and article features. Panel (d) details how reference titles and reference journals serve as indicators of the research community the article belongs to.}
    \label{fig:prompt_2}
\end{figure}

\paragraph{Part 2: Input Description.} The prompt incorporates two categories of input features, as shown in Figure~\ref{fig:prompt_2}. The first consists of journal and article features, including candidate journal scope descriptions that summarize the research focus and target audience of each journal, together with title, abstract, and keywords. The second category introduces citation features extracted from the reference list, including reference titles and reference journals, which provide signals about the research community to which the article belongs. However, these citation features are used only in the second stage analysis to examine the contribution of citation information and are not required from users in practical applications.

\begin{figure}[h]
    \centering
    \includegraphics[width=1\linewidth]{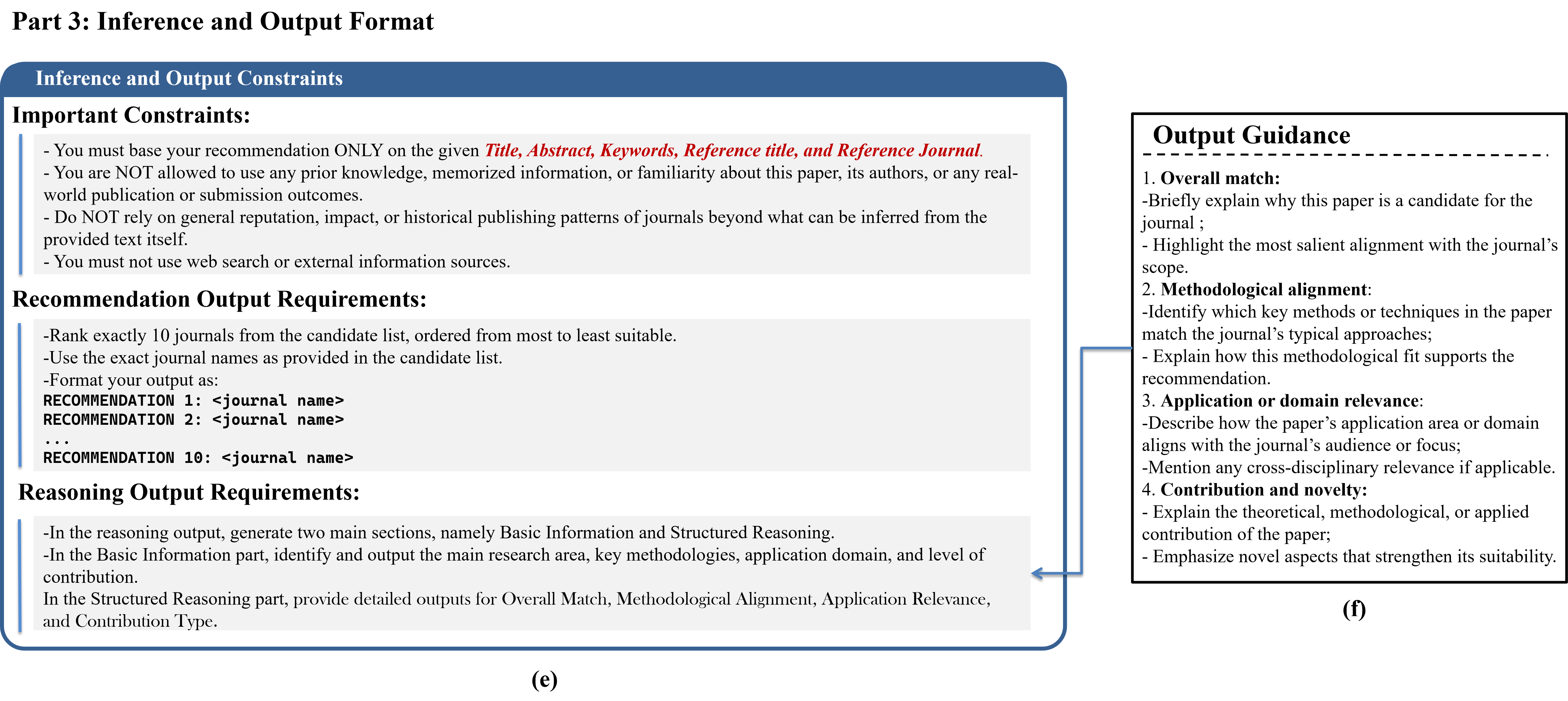}
    \caption{Illustration of the third part of the proposed prompt, covering inference and output constraints. Panel (e) presents the inference constraints and two output formats: a ranked list of 10 journals, and an optional structured reasoning output. Panel (f) details the reasoning guidance, specifying four aspects for each recommendation: overall match, methodological alignment, application relevance, and contribution type.}
    \label{fig:prompt_3}
\end{figure}

\paragraph{Part 3: Inference and Output Format.} The prompt specifies the constraints and output requirements to ensure controlled model behavior, as shown in Figure~\ref{fig:prompt_3}. 
To prevent information leakage, the model is restricted to using only the title, abstract, keywords, reference titles, and reference journals of the article, together with a predefined list of candidate journals. 
The output consists of two parts. In the primary setting, the model returns exactly 10 recommended journals from the candidate list, ranked from most to least suitable, 
using the exact journal names and no additional text. In the reasoning setting, the model generates structured explanations for each recommendation, covering four 
aspects: overall match, methodological alignment, application relevance, and 
contribution type. The reasoning outputs are used to examine whether the model recommendations reflect the content of the input article, with a detailed analysis 
provided in Section~\ref{sec:reason}.

\begin{figure}[h]
    \centering
    \includegraphics[width=1\linewidth]{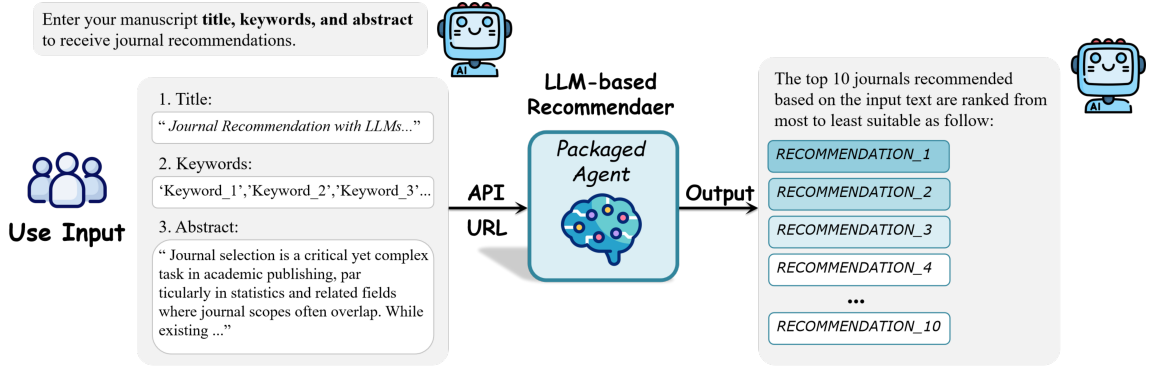}
    \caption{Illustration of the practical usage of the proposed journal recommendation system.}
    \label{fig:pre}
\end{figure}

Based on the proposed prompt framework, we implement the recommendation system as a practical agent using Python 3.9 and the DeepSeek API\footnote{Code is publicly available at: \mbox{\url{https://github.com/LLM-Journal-Rec/journal-recommendation-agent}}}. Figure~\ref{fig:pre} illustrates the workflow of the implemented system. Users are first required to provide an API key together with the corresponding API base URL to access the LLM service. The agent then takes article features, including the title, abstract, and keywords, as input and returns a ranked list of recommended journals. Note that reference information is not required from users in practical applications, since it is only incorporated in the second experiment (Section~\ref{sec:result_citation}) to evaluate the contribution of citation information. The system further supports both single paper and batch recommendation.

\section{Results and Enhancements}
\label{sec:ACC}

In this section, we first report the journal recommendation performance of the LLMs across five publication years. 
Specifically, we report the Top-3, Top-5, and Top-10 accuracies, and Mean Reciprocal Rank (MRR) for each year as well as the overall results. Top-$K$ accuracy is widely used in information retrieval and recommendation tasks to evaluate whether the relevant item appears within the first $K$ returned results \cite{manning2008introduction}. In addition, MRR further evaluates the ranking position of the target journal by assigning higher scores when the relevant journal appears closer to the top of the recommendation list \cite{voorhees1999trec8}.
We then examine the role of citation structure by comparing these accuracy metrics across eight representative journals, highlighting how incorporating citation information affects the recommendation behavior of LLMs. 

\subsection{Overall Performance}
\label{sec:result_noref}

In this section, we report the journal recommendation performance of DeepSeek-V3 on the base dataset described in Table~\ref{tab:dataset-stats}, including the Top-3, Top-5, and Top-10 accuracies, MRR, for each publication year from 2021 to 2025, along with the average sample size and runtime. As shown in Table~\ref{tab:overall_results_by_year}, the yearly sample size remains relatively stable over the five-year period, with an average of 4,721 articles per year. In terms of recommendation accuracy, the model achieves 40.23\%, 53.67\%, and 70.05\% for the Top-3, Top-5, and Top-10 metrics, respectively, indicating strong overall performance across all evaluation criteria. The clear increase from Top-3 to Top-10 further suggests that DeepSeek-V3 is often able to include the target journal within a broader candidate set, even when it does not rank it among the very top positions. This is practically meaningful for journal recommendation, since authors typically consider multiple candidate journals rather than relying on a single recommendation. Therefore, the model can effectively narrow down the journal search space and provide a useful shortlist for manuscript submission.
When examined across publication years, all three accuracy metrics exhibit a similar pattern, with relatively stronger performance in earlier years followed by a gradual decline in more recent years. Taking the Top-3 metric as an example, the accuracy increases slightly from 2021 to 2022, reaching a peak of 43.03\%. However, from 2023 onward, the accuracy gradually declines, decreasing to approximately 37\% by 2025. This trend may reflect the increasing breadth and diversity of statistical research, which leads to greater semantic overlap across journals and consequently makes journal recommendation more challenging. 

While Top-$K$ accuracy measures whether the target journal appears within a given recommendation range, it does not directly show how high the target journal is ranked within the list. To complement this perspective, MRR is further used to evaluate whether the target journal is ranked near the top of the recommendation list. Overall, DeepSeek-V3 achieves an average MRR of 0.344, indicating that the target journal is typically ranked within the top few positions of the recommendation list, approximately around the third position. 
Specifically, three observations can be drawn from the results.
First, the MRR values are relatively high in the earlier years, reaching 0.367 in 2021 and 0.368 in 2022. This result is consistent with the higher Top-3 accuracies in these two years, suggesting that the model not only identifies the target journal more often but also tends to place it in higher ranking positions. 
Second, MRR gradually declines from 0.348 in 2023 to 0.311 in 2025. Compared with its peak value of 0.368 in 2022, MRR decreases by approximately 15.5\% by 2025. This pattern is consistent with the decline in Top-3, Top-5, and Top-10 accuracies, indicating that recent articles are more difficult to rank accurately. A possible reason is that recent statistical research covers broader and more overlapping topics, making it harder to distinguish among closely related journals.
Third, although the MRR decreases over time, it remains above 0.300 in all years. This suggests that DeepSeek-V3 still maintains a reasonable ability to prioritize relevant journals, even when the recommendation task becomes more challenging. Therefore, the MRR results support the overall conclusion that DeepSeek-V3 can provide useful journal rankings rather than only broad candidate lists.
Overall, these results show that DeepSeek-V3 performs well in generating useful journal recommendations, especially in constructing a relevant candidate list, while its gradual decline in recent years also suggests room for improvement in fine-grained ranking among semantically similar journals.

\begin{table}[htbp]
\centering
\caption{Journal recommendation performance of DeepSeek-V3. The table reports the sample size, Top-3, Top-5, Top-10 accuracies, MRR, together with the computational time for each year from 2021 to 2025, where the final row summarizes the overall results.}
\label{tab:overall_results_by_year}
\newcolumntype{C}{>{\centering\arraybackslash}X}
\begin{tabularx}{\textwidth}{lCCCCCC}
\toprule
Year & Sample Size & Top-3(\%) & Top-5(\%) & Top-10(\%) & MRR & Run Time \\
\midrule
2021 & 4,288 & 42.37 & 56.53 & 72.67 & 0.367 & 1h 45min \\
2022 & 4,660 & 43.03 & 56.52 & 72.66 & 0.368 & 1h 51min \\
2023 & 4,672 & 40.26 & 53.90 & 69.54 & 0.348 & 1h 44min \\
2024 & 5,308 & 38.79 & 51.87 & 68.52 & 0.330 & 1h 55min \\
2025 & 4,681 & 37.06 & 50.03 & 67.29 & 0.311 & 1h 31min \\
\midrule
Average & 4,721 & 40.23 & 53.67 & 70.05 & 0.344 & 1h 45min \\
\bottomrule
\end{tabularx}
\end{table}

\subsection{Effect of Citation Structure}
\label{sec:result_citation}

In scholarly communication, citations link related studies and provide informative signals regarding the research area and methodological framework of an article. Therefore, examining the role of citation information can help evaluate whether citation-based signals improve journal recommendation performance in academic publishing scenarios.
In Section \ref{sec:result_noref}, the journal recommendation task is based on article-level information, including the title, abstract, and keywords as indicators of its research scope. In this section, we extend the primary experiment by incorporating citation information to investigate whether it can further improve journal recommendation performance.
The experiment is conducted on the citation dataset introduced in Section~\ref{sec:data}, which consists of articles published in eight core statistical journals, i.e.,  \textit{AoAS}, \textit{AoS}, \textit{Bka}, \textit{JASA}, \textit{JBES}, \textit{JCGS}, \textit{JRSSB}, and \textit{SCmp}. More details can be found in Table~\ref{tab:journals_full}.
Specifically, we augment the input by incorporating the reference titles and reference journals of cited works, while keeping the task objective unchanged. The dataset statistics and input features are summarized in Table~\ref{tab:dataset-stats} and Table~\ref{tab:example}, respectively. The study is further based on a journal citation network constructed in \citep{xie2025bi}.

\begin{table}[htbp]
\centering
\caption{Journal-level recommendation accuracy with and without reference information. The table reports the Top-3, Top-5, and Top-10 accuracies, together with their corresponding absolute changes.}
\label{tab:journal_ref}

\setlength{\tabcolsep}{3pt}  

\begin{tabularx}{\textwidth}{lZZZZZZZZZ}
\toprule
\multirow{2}{*}{Journal Name} 
& \multicolumn{3}{c}{Without Reference (\%)} 
& \multicolumn{3}{c}{With Reference (\%)} 
& \multicolumn{3}{c}{Change (\%)} \\
\cmidrule(lr){2-4} \cmidrule(lr){5-7} \cmidrule(lr){8-10}
& Top-3 & Top-5 & Top-10 
& Top-3 & Top-5 & Top-10 
& Top-3 & Top-5 & Top-10 \\
\midrule
AoAS       & 21.61 & 61.75 & 90.07 & 30.28 & 56.94 & 80.28 &  8.67 &  -4.81 &  -9.79 \\
AoS        & 83.02 & 87.65 & 95.03 & 66.86 & 79.32 & 94.33 & -16.16 &  -8.33 & -0.70 \\
Bka        & 18.97 & 28.73 & 36.31 & 20.88 & 35.34 & 52.21 &  1.91 &   6.61 &  15.90 \\
JASA       & 32.79 & 54.85 & 91.11 & 45.17 & 72.86 & 89.03 & 12.38 &  18.01 &  -2.08 \\
JBES       & 74.69 & 77.76 & 82.04 & 83.33 & 88.10 & 92.18 &  8.64 &  10.34 &  10.14 \\
JCGS       & 43.29 & 57.01 & 72.87 & 41.03 & 60.11 & 80.34 & -2.26 &   3.10 &   7.47 \\
JRSSB     & 33.44 & 47.85 & 74.23 & 38.42 & 61.58 & 86.44 &  4.98 &  13.73 &  12.21 \\
SCmp         & 37.38 & 58.29 & 71.53 & 61.22 & 70.64 & 80.89 & 23.84 &  12.35 &  9.36 \\
\bottomrule
\end{tabularx}
\end{table}

To analyze the results, we compare the recommendation performance of DeepSeek-V3 with and without reference information. The detailed results are presented in Table~\ref{tab:journal_ref}, and three key observations are discussed below.
First, most journals exhibit improved recommendation accuracy after incorporating reference information. For example, journals such as \emph{JBES} and \emph{JRSSB} show consistent gains across the Top-3, Top-5, and Top-10 metrics. In particular, \emph{SCmp} exhibits the largest improvement, with its Top-3 accuracy increasing by 23.84\%.
Second, such improvements are not universal. In particular, \emph{AoAS} and \emph{AoS} show noticeable declines after incorporating reference information. For \emph{AoAS}, which focuses on applied statistics, published papers often span diverse application domains. Their reference lists tend to cover a wide range of research areas, making it harder for the model to identify clear disciplinary signals from the reference information. For \emph{AoS}, a theory-oriented journal, articles typically exhibit a clear and focused methodological profile that can already be effectively captured by the title, abstract, and keywords alone. However, their reference lists may involve a broader body of theoretical literature, introducing additional noise into the input and thereby reducing recommendation precision.
Third, some broad-scope journals exhibit overall improvements while still experiencing minor declines in specific metrics. Journals such as \emph{JASA} and \emph{JCGS} generally benefit from the inclusion of reference information, although slight decreases are observed for certain metrics, such as Top-10 accuracy for \emph{JASA} and Top-3 accuracy for \emph{JCGS}. This may be attributable to their broad scope, where the reference lists introduce mixed disciplinary signals that reduce precision under specific evaluation metrics.

\section{Stability, Leakage Risk, and Interpretability}
\label{sec:stability}

In this section, we first evaluate the stability of the LLM by computing the Top-5 overlap count and Jaccard similarity across five repeated runs. Second, we examine the potential issue of information leakage and discuss the corresponding mitigation strategies. Finally, we present a representative case study to investigate whether the model can generate structured and interpretable reasoning for journal recommendation.

\subsection{The Stability of LLMs}

Stability is an important consideration in LLM-based recommendation systems, as model outputs may vary across repeated runs even when the inputs remain identical. To assess the stability of our approach, we randomly sample 500 papers per year from the base dataset (Table~\ref{tab:dataset-stats}), resulting in a total of 2,500 papers. The recommendation process is then repeated five times using the same prompts and inputs.
To quantify stability, we adopt two metrics based on the Top-5 recommendation lists. The first is the overlap count, defined as the number of shared journals between two Top-5 recommendation sets from a pair of runs. The second is the Jaccard similarity, computed as the ratio of the intersection size to the union size of two Top-5 sets. Both metrics are averaged across all papers and all pairs of runs, and the results are reported in Table~\ref{tab:stability_matrix}.

The results show that the Top-5 recommendations are relatively consistent across runs, with approximately four journals overlapping on average between any pair of runs and a Jaccard similarity of around 84\%. Moreover, both metrics exhibit very limited variation across different pairs of runs. The overlap count ranges only from 4.47 to 4.49, while the Jaccard similarity varies between 83.90\% and 84.69\%. Such small fluctuations indicate that the recommendation results are highly reproducible under repeated executions with identical inputs and prompts.
Although the recommendation lists are not identical across runs, the differences are small and remain within an acceptable range. The consistently high overlap and Jaccard values suggest that the variability mainly affects a small number of journals rather than causing substantial changes to the overall recommendation set. These findings indicate a moderate to high degree of agreement among recommendation sets, showing that DeepSeek-V3 can provide reliable journal suggestions across repeated runs in practical manuscript submission scenarios.

\begin{table}[htbp]
\centering
%\footnotesize
\renewcommand{\arraystretch}{1.3}
\caption{Pairwise Top-5 overlap counts and Jaccard similarities across five repeated runs. Each cell reports the average number of overlapping journals together with the corresponding Jaccard similarity (in \%). For example, the entry ``4.49 (84.69)'' for Exp1 and Exp2 indicates that their Top-5 recommendation lists share an average of 4.49 journals, corresponding to a Jaccard similarity of 84.69\%.}
\label{tab:stability_matrix}
\newcolumntype{C}{>{\centering\arraybackslash}X}
\begin{tabularx}{\textwidth}{lCCCCC}
\toprule
& Exp1 (\%) & Exp2 (\%) & Exp3 (\%) & Exp4 (\%) & Exp5 (\%) \\
\midrule
Exp1 & -- & 4.49 (84.69) & 4.47 (84.05) & 4.49 (84.42) & 4.48 (84.17) \\
Exp2 &    & --           & 4.47 (84.11) & 4.47 (84.00) & 4.47 (83.90) \\
Exp3 &    &              & --           & 4.48 (84.25) & 4.47 (83.99) \\
Exp4 &    &              &              & --           & 4.48 (84.26) \\
Exp5 &    &              &              &              & --           \\
\midrule
Average & & & & & 4.48 (84.16) \\
\bottomrule
\end{tabularx}
\end{table}

\subsection{Potential Information Leakage}

A natural concern in applying LLMs to journal recommendation is the risk of information leakage. Contemporary LLMs may be pretrained on large-scale corpora that include historical publication records and journal–article associations. As a result, LLMs may contain implicit knowledge about where certain types of articles are typically published, a phenomenon widely discussed in recent studies on LLM pretraining and evaluation \citep{raji2020closing, carlini2021extracting}.
If left unconstrained, such knowledge may bias journal recommendation results toward historical publication patterns. We emphasize that our study does not aim to replicate real-world submission outcomes or editorial decisions based on historical publishing data. Instead, we investigate whether LLMs can perform journal recommendation under controlled prompt constraints. The evaluation relies solely on contemporaneous textual information and focuses on the semantic compatibility between articles and candidate journals. These considerations highlight the importance of controlling potential information leakage in prompt design for journal recommendation tasks.

To control information leakage, the prompt design adopts a series of structural constraints. First, the prompt restricts the decision process to information explicitly provided in the input text. Each prompt corresponds to a single article and includes only the title, abstract, keywords, reference title, and reference journal, together with predefined journal scope descriptions. This design reduces the probability of information leakage at the source \citep{agarwal2024prompt}.
Second, the journal recommendation task is formulated independently of the real-world publication context. Candidate journals are represented as semantic profiles defined by their scope descriptions, and the recommendation is formulated as a compatibility assessment between the article content and journal profiles. This design restricts the association between the model and real-world entities \citep{hui2024pleak}.
Third, the output is strictly constrained in both content and format. The model is required to return only a ranked list of candidate journals using the exact journal names. By prohibiting free-form explanations, this design reduces the opportunity for the model to expose memorized training content, historical publication patterns, or latent journal knowledge during generation. As a result, the risk of information leakage during the recommendation process can be mitigated \citep{hou2025lne}.
This constraint is used in the main recommendation experiments, whereas the reasoning analysis in Section~\ref{sec:reason} is conducted separately for interpretability purposes.
Finally, a shielding mechanism is introduced at the prompt level. The prompt prohibits the use of prior knowledge, journal reputation or impact, historical publishing patterns, and external search, ensuring that model outputs are derived exclusively from the controlled textual inputs. This design enhances the stability of the model in following information-usage constraints under complex instructions \citep{jawad2025psm}.

Under these constraints, journal recommendations rely on patterns in research topics, methodological focus, and scientific or application domains reflected in the provided text. As a result, potential information leakage effects are mitigated at the prompt level. We acknowledge that fully eliminating all forms of latent knowledge related to publication would require retraining or temporally constrained pretraining of LLMs. Such model-level interventions are beyond the scope of this study. The present design therefore focuses on prompt-level controls that constrain accessible information without modifying model parameters. Accordingly, the results should be interpreted as evidence of the extent to which LLMs can approximate journal recommendation judgments based on controlled textual inputs. The goal is not to reproduce editorial decisions or historical publication outcomes. Instead, the analysis evaluates whether consistent journal recommendation can be achieved when the process is restricted to controlled textual inputs available at submission time.

\subsection{The Interpretability of LLMs}
\label{sec:reason}

\begin{figure}[!ht]
    \centering
    \includegraphics[width=1\linewidth]{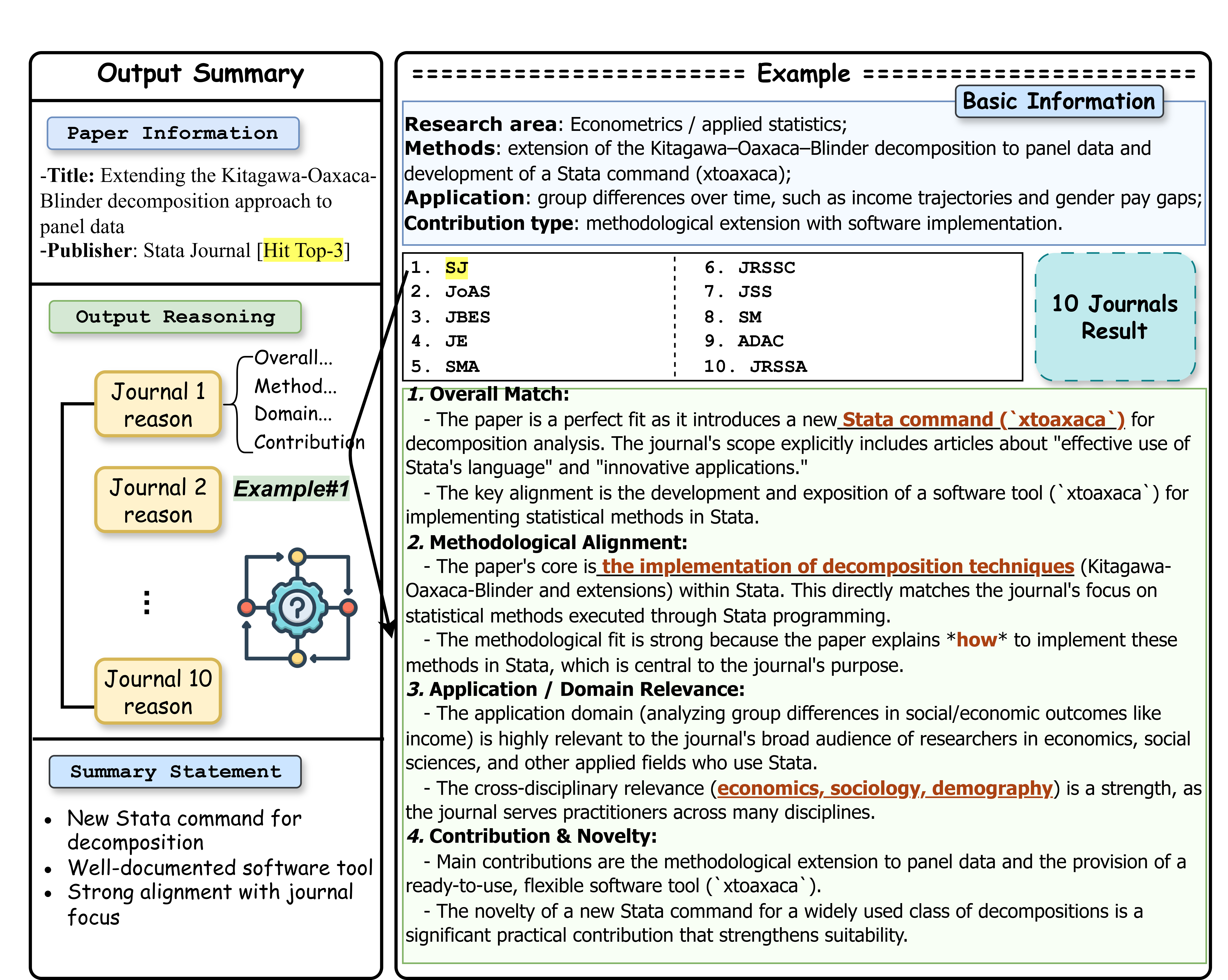}
    \caption{An example of LLM-based reasoning for journal recommendation. The figure illustrates the detailed reasoning process of the LLM for journal recommendation, where the left panel summarizes the output mechanism. The reasoning output consists of two main components, namely \textit{Basic Information} and \textit{Journal-Level Reasoning}. In addition, the Top-10 recommended journals for this example are presented for comparison.}
    \label{fig:Reason}
\end{figure}

Existing studies show that LLMs are capable of generating natural language explanations for their recommendations \citep{ma2024xrec}. In this section, we further investigate how the LLM produces its journal recommendations by prompting it to generate reasoning outputs.
Specifically, based on the Top-10 journal recommendations obtained in Section~\ref{sec:result_noref}, we prompt the model to first summarize the basic information of the target article and then generate corresponding reasoning for each recommended journal, as illustrated in the prompt design shown in Figure~\ref{fig:prompt_3}. Due to space limitations, only the reasoning output for one recommended journal is presented. An illustrative example of the generated reasoning output is shown in Figure~\ref{fig:Reason}.
In this example, the Top-1 recommendation matches the true publisher, namely \emph{Stata Journal}, indicating a correct recommendation. The reasoning output provides insight into how this recommendation is reached. In the \textit{Basic Information} section, the model identifies the article as an applied, software-oriented methodological contribution, which serves as the primary signal for recommendation.
From the \textit{Overall Match} and \textit{Methodological Alignment} sections, the model further identifies the core contribution of the article as a Stata-based decomposition tool, which directly aligns with the scope of \emph{Stata Journal}. This example suggests that the model is able to produce consistent and interpretable 
reasoning that captures the relevance between the input article and the recommended journal. More broadly, such structured reasoning outputs provide researchers with a clearer basis for evaluating the recommendations, thereby improving the interpretability of the system and helping researchers better identify suitable target journals in practice.

\section{Discussion}

In summary, the primary goal of this study is to examine whether LLMs can perform journal recommendation using only submission-stage information. This setting is challenging because the model does not rely on historical submission records, user behavior, citation networks, or task-specific supervised training. Instead, the recommendation is based only on the semantic alignment between manuscript information and journal scope descriptions. The empirical results indicate that DeepSeek-V3 achieves Top-3, Top-5, and Top-10 accuracy of 40.23\%, 53.67\%, and 70.05\%, respectively, which is broadly competitive with existing methods summarized in Table~\ref{tab:journal_rec}. For instance, compared with TF-IDF (Top-3: 42\%, Top-5: 50\%, Top-10: 58\%), DeepSeek-V3 achieves comparable performance without requiring any task-specific training or index construction. This finding suggests that LLMs can capture useful semantic relationships between manuscripts and candidate journals in a zero-shot recommendation setting. Additional experiments further validate several properties of the proposed approach, including the benefit of incorporating reference information, the stability of model outputs, and the ability to produce interpretable reasoning for each recommendation. These results suggest that LLMs provide a viable approach for journal recommendation when only submission-stage information is available.

\begin{table*}[htbp]
\centering
\small
\setlength{\tabcolsep}{4pt}
\renewcommand{\arraystretch}{1.2}
\caption{Summary of existing journal recommendation studies. The Method column reports the primary model or core approach adopted in each study. In the Performance column, we report only metrics comparable to our evaluation framework.}
\label{tab:journal_rec}
\begin{threeparttable}
\begin{tabularx}{\textwidth}{>{\centering\arraybackslash}p{1.8cm} p{2.8cm} p{2.8cm} p{3.2cm} >{\raggedleft\arraybackslash}X}
\toprule
\textbf{Method} & \textbf{Dataset} & \textbf{Input Features} & \textbf{Performance} & \textbf{Reference} \\
\midrule

% ── Group 1–3: TF-IDF ────────────────────────────────────────────────
\multirow{2}{=}{\centering TF-IDF}
% & \makecell[lt]{245,573 papers;\\ 78 journals\\ (AI \& Medicine)}
% & title, abstract, keywords
% & \makecell[lt]{(Top-5):ACC=0.84(AI);\\(Top-5)ACC=0.92(MED)}
% & \cite{schafermeier2021towards} \\
% \cmidrule(l){2-5}
& \makecell[lt]{1,000 papers$^{\rm a}$\\(Biomedical)}
& title, abstract
& \makecell[lt]{(Top-3): ACC=0.42;\\ (Top-5): ACC=0.50;\\ (Top-10): ACC=0.58}
& \cite{schuemie2008jane} \\
\cmidrule(l){2-5}
& \makecell[lt]{14,012 papers$^{\rm b}$;\\ 66 venues\\ (CS)}
& abstract
& \makecell[lt]{(Top-3):ACC=0.61}
& \cite{wang2018content} \\
\midrule

% ── BM25, Bibliographic Coupling ─────────────────────────────────────
\makecell[lt]{BM25,\\ Bibliographic\\ Coupling}
& \makecell[lt]{10,000 papers$^{\rm c}$\\ (Multi-disciplinary)}
& title, abstract, references
& \makecell[lt]{(Top-5): ACC=0.56;\\ (Top-10): ACC=0.68}
& \cite{entrup2024comparing} \\
\midrule

% ── Row 4: Clustering ────────────────────────────────────────────────
Clustering
& \makecell[lt]{309,551 papers;\\ 1,002 journals\\ (CS)}
& title, abstract, authors, terms
& \makecell[lt]{(Top-10): ACC=0.72 \\ (CB+AU, GP K=110)}
& \cite{de2022publication} \\
\midrule

% ── Row 5: Doc2Vec + XGBoost ─────────────────────────────────────────
Doc2Vec + XGBoost
& \makecell[lt]{20,250 papers;\\ 45 journals$^{\rm d}$\\ (CS)}
& title, abstract, keywords
& \makecell[lt]{(Top-3): ACC=0.84;\\ (Top-5): ACC=0.89}
& \cite{zhengwei2022recommendation} \\
\midrule

% ── Row: kNN ─────────────────────────────────────────────────────────
\multirow{2}{=}{\centering kNN}
& \makecell[lt]{960 papers;\\ 16 conferences\\ (CS)}
& co-author network, publication history
& \makecell[lt]{(Top-3): ACC=0.81}
& \cite{luong2012publication} \\
\cmidrule(l){2-5}
& \makecell[lt]{a subset of DBLP\\ (CS)}
& title, abstract, authors, year
& \makecell[lt]{(Top-3): ACC=0.36;\\ (Top-6): ACC=0.39;\\ (Top-9): ACC=0.46}
& \cite{pradhan2020cnaver} \\
\midrule

% % ── Row 8: RNN ───────────────────────────────────────────────────────
% RNN
% & \makecell[lt]{738,141 papers\\ (CS)}
% & title, authors, citations
% & \makecell[lt]{(Top-20): ACC=0.1590;\\ Recall=0.4055;\\ MRR=0.6568;\\ NDCG=0.6581}
% & \cite{zhu2021recommending} \\
% \midrule

% ── Row 9: GNN ───────────────────────────────────────────────────────
GNN
& \makecell[lt]{137,376 papers;\\ 1,122 conferences\\ (CS)}
& title, abstract, authors, network
& (Recall@10): 0.58
& \cite{iana2021graphconfrec} \\
\midrule

% ── SBERT ────────────────────────────────────────────────────────────
SBERT
& \makecell[lt]{184,055 papers;\\ 598 journals\\ (CS)}
& title, abstract, keyword, references, journal scope
& \makecell[lt]{(Top-15): ACC=0.48;\\ (Top-20): ACC=0.60;\\ (Top-25): ACC=0.56;\\ (Top-30): ACC=0.72}
& \cite{gundougan2023deep} \\
\bottomrule
\end{tabularx}
\begin{tablenotes}
\scriptsize
\item[$^{\rm a}$] 1,000 papers randomly sampled from PubMed; underlying index: 4,171,368 articles from 4,513 journals.
\item[$^{\rm b}$] Training and test sets split in a 2:1 ratio.
\item[$^{\rm c}$] 10,000 papers randomly sampled from DOAJ; underlying index: 8,154,699 articles from 18,461 journals.
\item[$^{\rm d}$] 450 papers per journal (400 training + 50 test).
\end{tablenotes}
\end{threeparttable}
\end{table*}

% Several limitations of the present study should be acknowledged. First, the candidate pool of 49 journals is substantially smaller than those used in prior studies, as summarized in Table~\ref{tab:journal_rec}, and it remains unclear whether the results generalize to larger candidate sets. Second, the current analysis focuses on a single model, DeepSeek-V3, while recommendation performance may vary across LLMs with different architectures and training strategies. Third, the proposed framework generates recommendations based solely on static textual information available at submission time, without considering other factors that may influence journal selection in practice, such as author preferences and prior submission behavior. 

% These limitations point to several directions for future work. Evaluating the framework on larger and more diverse candidate pools would help assess its scalability and generalizability. In addition, systematic comparisons across different LLMs could provide further insight into how model architectures and training strategies affect recommendation performance. Finally, future work may explore modeling journal selection as a sequential decision-making process in order to better reflect real-world submission and resubmission dynamics. 

Several limitations of the present study should be acknowledged. First, the candidate pool of 49 journals is substantially smaller than those used in prior studies, as summarized in Table~\ref{tab:journal_rec}, and it remains unclear whether the results generalize to larger candidate sets. As the number of candidate journals increases, the semantic overlap among journals may become more complex, making fine-grained ranking more difficult. Second, the current analysis focuses on a single model, DeepSeek-V3, while recommendation performance may vary across LLMs with different architectures and training strategies. Third, the proposed framework generates recommendations based solely on static textual information available at submission time, without considering other factors that may influence journal selection in practice, such as author preferences, prior submission behavior, review speed, open-access policy, and journal impact. Therefore, the findings should be viewed as evidence that LLMs can support journal recommendation through semantic matching, rather than as a complete representation of all factors involved in journal selection.

These limitations point to several directions for future work. Evaluating the framework on larger and more diverse candidate pools would help assess its scalability and generalizability. In addition, systematic comparisons across different LLMs could provide further insight into how model architectures and training strategies affect recommendation performance. Future work may also incorporate richer journal-level information, such as publication history, and review-related metadata, to better capture the practical considerations involved in journal selection. Finally, future work may explore modeling journal selection as a sequential decision-making process in order to better reflect real-world submission and resubmission dynamics.

% The present study evaluates 49 journals under controlled conditions, it remains unclear whether the observed accuracy can be maintained when the candidate pool expands to hundreds or thousands of journals with overlapping scopes.

\clearpage
\renewcommand{\bibsection}{}
\section{REFERENCES}
\bibliographystyle{ACM-Reference-Format}
\bibliography{ref}

\end{document}